\documentclass[fleqn,usenatbib]{mnras}

\usepackage{newtxtext,newtxmath}

\usepackage[T1]{fontenc}

\DeclareRobustCommand{\VAN}[3]{#2}
\let\VANthebibliography\thebibliography
\def\thebibliography{\DeclareRobustCommand{\VAN}[3]{##3}\VANthebibliography}

\usepackage{graphicx}	
\usepackage{amsmath}	

\title[Tidal resurfacing of Apophis]{Tidal resurfacing model for (99942) Apophis during the 2029 close approach with Earth}

\author[Y. Kim et al.]{
Yaeji Kim,$^{1}$\thanks{E-mail: yzk0056@auburn.edu}
Joseph V. DeMartini,$^{2}$
Derek C. Richardson$^{2}$
and Masatoshi Hirabayashi$^{1,3}$
\\
$^{1}$Department of Aerospace Engineering, Auburn University, 211 Davis Hall, Auburn, Alabama 36849, USA\\
$^{2}$Department of Astronomy, University of Maryland, 4296 Stadium Dr, College Park, Maryland 20742, USA\\
$^{3}$Department of Geosciences, Auburn University, 2050 Beard Eaves Coliseum, Auburn, Alabama 36849, USA
}

\date{Accepted XXX. Received YYY; in original form ZZZ}

\pubyear{2022}

\begin{document}
\label{firstpage}
\pagerange{\pageref{firstpage}--\pageref{lastpage}}
\maketitle

\begin{abstract}
We numerically investigate tidally induced surface refreshing on Apophis during its close approach with Earth within a perigee distance of 5.96 Earth radii on April 13, 2029. We implement a tidal resurfacing model with two stages: dynamics modeling of the entire body to determine time-varying accelerations and surface slope profiles felt by each surface patch during the 6-h-long closest encounter, and DEM modeling to track motions of surface grains in localized patches. The surface slope profiles and measured grain motions are combined to statistically extrapolate the `expected' percentage of resurfaced area. Using the tidal resurfacing model, we present surface maps showing the total expected resurfacing on Apophis given 3 representative encounter orientations. Our simulation results indicate that tidal resurfacing, limited to certain localized regions, will likely occur half an hour before perigee and on the scale of 1 per cent of Apophis's entire surface area. Our models indicate that the most likely locations to detect tidal resurfacing are: initially high-sloped regions ($> 30^{\circ}$) regardless of the encounter orientation of Apophis, and mid-sloped regions ($15^{\circ}$--$30^{\circ}$) that experience a significant positive slope variation ($> 0.5^{\circ}$), which is mainly controlled by the encounter orientation. Expected data from ground-based observations of the 2029 flyby will help us better constrain the targeted locations likely to experience tidal resurfacing. We thus expect to find evidence supporting tidal resurfacing via further analysis of post-encounter surface images or albedo changes at the expected resurfaced areas.
\end{abstract}




\section{Introduction}
\label{sec:intro}
(99942) Apophis is a potentially hazardous asteroid that will pass the Earth within a perigee distance of 5.96 Earth radii on 2029 April 13\footnote{The perigee distance is the minimum possible close-approach distance between the 3-sigma Earth target-plane error ellipse and the Earth’s surface, retrieved from the Center for Near-Earth Object Studies server (\url{https://cneos.jpl.nasa.gov/ca/}).}. The 2029 Apophis-Earth encounter event is considered a golden opportunity to directly observe how Earth-crossing objects interact with the Earth's gravity field, offering a natural experiment which could allows us to better understand potentially hazardous objects and support the science of planetary defense \citep{binzel2020apophis}. As a result of the unique opportunities for science that this object's passage provides, NASA has recently announced that Apophis has been selected as the target of the OSIRIS-REx extended mission---OSIRIS-APEX \citep{dellagiustina2022osiris}.

The perigee distance of Apophis from Earth during the 2029 close encounter is outside the canonical Roche limit ($\sim$3.4 Earth radii, given a bulk density of $\sim$2 g~cm$^{-3}$ for an Sq-type asteroid like Itokawa \citep{abe2006mass}) that induces catastrophic disruptions of unconsolidated material \citep{richardson1998tidal,zhang2020tidal}. Note that Apophis is intermediate between S- and Q-type asteroids \citep{binzel2009spectral}. During the 2029 Earth encounter, Apophis will have a definite change in its orbit and rotational properties in response to Earth's tidal torques \citep{scheeres2005abrupt,farnocchia2013yarkovsky,souchay2014rotational,souchay2018changes,demartini2019using, benson2022spin}. These orbital and rotational changes will likely occur with magnitudes sufficient to be detectable by ground-based telescopes. Furthermore, we anticipate that the perigee distance of 5.96 Earth radii may be close enough to subject Apophis to some influences from Earth's tidal forces: surface refreshing \citep{yu2014numerical,de2021tidal}, small-scale structure modifications and seismic vibrations \citep{demartini2019using}, and stress variation around Apophis's concave region \citep{hirabayashi2021finite}. Among the potential consequences of the tidal encounter, we particularly note surface refreshing, which may be detectable during the 2029 Apophis-Earth encounter and which is the primary focus of this study.

In general, asteroid surfaces are affected by the competing processes of space weathering and mechanical resurfacing, creating variations in their surface colors. Space weathering reddens or darkens surface materials as a result of solar wind irradiation or micrometeorite impacts, and has been commonly observed in S-type asteroids \citep{pieters2000space,sasaki2001production,pieters2016space, thompson2021space}. Resurfacing is an opposing mechanism that exposes fresh materials beneath the weathered asteroid surface. The interplay between weathering and resurfacing resulting in a variegated surface can be seen on the S-type asteroid Itokawa \citep{fujiwara2006rubble, miyamoto2007regolith, jin2022estimation} and the Martian moon Phobos \citep{fraeman2014spectral, ballouz2019surface}, which appear to have dark/redder surfaces with some bright/bluer regions. Although there are other potential mechanisms (e.g., thermal fatigue \citep{delbo2014thermal}, YORP (Yarkovsky–O'Keefe–Radzievskii–Paddack) spin-up \citep{graves2018resurfacing}, and impact-induced seismic shaking \citep{yamada2016timescale}), surface refreshing as a result of planetary encounters is one possible contributor to the inferred resurfacing in near-Earth asteroids (NEAs). 

In the taxonomic classification of NEAs, S- and Q-types show different absorption features and spectral slopes \citep{chapman1996s,chapman2004space} although both represent the same compositions as ordinary chondrites. S-types show more reddened surfaces indicative of space weathering, while Q-types have bluer surfaces indicating relatively fresh surface materials. To resolve this inconsistency, \cite{nesvorny2005evidence} suggest that the relatively unweathered surfaces of Q-types result from surface regolith motion during tidal encounters. Many subsequent studies (e.g., \cite{marchi2006spectral,binzel2010earth,nesvorny2010planetary,demeo2014mars}) support this hypothesis by statistically showing that the distribution of Q-type asteroid orbital parameters correlates with low perihelion distances and low minimum orbit intersection distances (MOID) with the terrestrial planets Earth, Mars, and Venus; \cite{marchi2006spectral} found that Q-types have lower perihelion distances than S-types using a data set of spectroscopic observations of NEAs and Mars-crossing asteroids. \cite{binzel2010earth} used a sample of 95 Earth- and Mars-crossing asteroids (including 20 Q-types) and revealed that Q-types more frequently experienced an Earth encounter with a limiting distance inside $\sim$15 Earth radii in the past few hundred thousand years. \cite{demeo2014mars} then used a larger data set of NEAs (including 64 Q-types) and identified that all sampled Q-types have low MOID values allowing either Earth or Mars encounters. A plausible mechanism for planetary encounters resurfacing weathered asteroid exteriors is that the tidal forces on the surfaces during the encounters fluidize the surface regolith, causing granular flows (i.e., landslides), which can move weathered materials and expose fresher subsurface materials. Based on this mechanism, \cite{keane2014rejuvenating} implemented a resurfacing model that evaluates the stability of asteroid regolith during distant planetary flybys using the theory of hillslope stability. The study set two free parameters, spin period and periapsis, and found that rapidly rotating asteroids are more likely to have surface conditions susceptible to resurfacing and that the limiting distance of resurfacing is less than 10 planetary radii. The asteroid was modeled as a triaxial ellipsoid with an arbitrarily defined density and a friction angle of 45$^{\circ}$. \cite{kim2021surface} extended this work to investigate how an asteroid's shape affects resurfacing and found that a more elongated shape tends to have unstable surface conditions to granular flows during a distant planetary encounter. All previous work has investigated tidal resurfacing from theoretical considerations; however, a direct observation of this phenomenon has never been made. The 2029 Apophis-Earth close encounter could mark the first observations that provide evidence of tidal resurfacing.

In this study, we visit the scientific question of whether tidal resurfacing will occur on Apophis's surface during its close Earth encounter. We use a tidal resurfacing model, which is a joint approach of dynamics \citep{kim2021surface} and discrete-element method (DEM) modeling \citep{yu2014numerical,demartini2019using} to numerically investigate the motion of surface grains driven by the tidal forces on Apophis during the Earth encounter. The dynamics model determines time-varying accelerations felt by each surface patch on Apophis, and the DEM model tracks the specific motion of grains on the given surface patch in the dynamical state. The results of this study could support an investigation of albedo changes after the Apophis close encounter or identify regions of interest to look for evidence of surface grain motion for potential missions to Apophis, including OSIRIS-APEX. Furthermore, understanding the tidal resurfacing processes on Apophis may provide key information about how resurfacing counteracts the expected space weathering timescale on small bodies, and could thus help resolve the long-standing puzzle of the spectral difference of NEAs between S- and Q-types despite their matching compositions \citep{chapman1995near,chapman1996s}. 

This paper is organized as follows. In Section \ref{sec:Methodology}, we describe the tidal resurfacing model in detail by splitting it into two parts: dynamics and DEM modeling. Section \ref{Sec:results} then shows our simulation results using the tidal resurfacing model. In Section \ref{Sec:Discussions}, we discuss the key findings to support potential ground-based observations and in-situ missions for the upcoming 2029 Apophis-Earth encounter event. Lastly, we summarize our conclusions, list areas of uncertainty in the current tidal resurfacing model, and suggest future work in Section \ref{Sec:futurework}.

\section{Tidal resurfacing model} \label{sec:Methodology}
We introduce a numerical model (hereafter `tidal resurfacing model') used for investigating surface grain motions driven by Earth-induced tides during Apophis's 2029 Earth flyby. In the following subsections, we split the numerical approach into two parts: dynamics and DEM modeling. The dynamics model simulates the orbital and spin evolution of Apophis during a period spanning 3 h before to 3 h after the closest encounter with Earth. By considering the local topographic features, the dynamics model converts the acceleration data into surface slope profiles, including an initial surface slope and slope variation, and then hands off the time-varying accelerations acting on each surface facet during the encounter to use in the DEM models. For the second stage of the simulations, we use DEM modeling to track the specific motion of grains on surface patches. We apply the time-varying acceleration data derived from the dynamics model uniformly across a number of surface patches filled with discrete regolith particles. The grain motion that we see from the discrete modeling forms the basis of our resurfacing analysis, detailed in Section \ref{Sec:results}.

\subsection{Dynamics model}
To investigate tidal refreshing during the Apophis-Earth close encounter, our dynamics model computes the acceleration vectors acting on surface facets of Apophis at each time step. We use the radar-derived, concave polyhedral shape model, consisting of 3,996 facets and 2000 vertices, by \cite{brozovic2018goldstone}. The \cite{brozovic2018goldstone} model was derived using the radar observations during the 2012--2013 apparition at Goldstone radar telescope facility in California (aka.\ Goldstone) in addition to the pre-existing lightcurve-derived convex shape by \cite{pravec2014tumbling}. The current shape model still has some uncertainties, which are unlikely to be improved by using the recently obtained 2020--2021 apparition data \citep{lee2022refinement}, but indicates that Apophis may likely be a contact binary. The net surface acceleration on each surface patch can be computed as the combination of self-gravity, tidal, and rotational accelerations. The detailed equations and propagation for each term are described in Appendix \ref{Appendix_dynamics}. We consider a time span of 6 h: 3 h before to 3 h after the closest Earth encounter, when the tidal effects are sufficient to induce variations in the total acceleration, including the fixed self-gravity and rotational accelerations. Outside of this time span, the tides are unlikely to induce any significant force variations because the distance of Apophis from the Earth's center, which exceeds 10 Earth radii, is far enough to neglect the tidal effect. We retrieve Apophis's trajectory using the JPL/NAIF SPICE tool \citep{acton1996ancillary,acton2018look} for the 6 h encounter with a timestep of 0.1 s. As a final note, Apophis is a tumbling object undergoing short-axis-mode non-principal-axis rotation. Given the slow spin period of 30.6 h \citep{pravec2014tumbling, brozovic2018goldstone} and the short (6 h) time span considered in our simulations for tidal resurfacing, the effect resulting from the tumbling motion of the body is likely negligible in the acceleration variation. Thus we propagate the rotation term assuming Apophis is in the simpler principal-axis rotation mode.

\begin{figure}
	\includegraphics[width=\columnwidth]{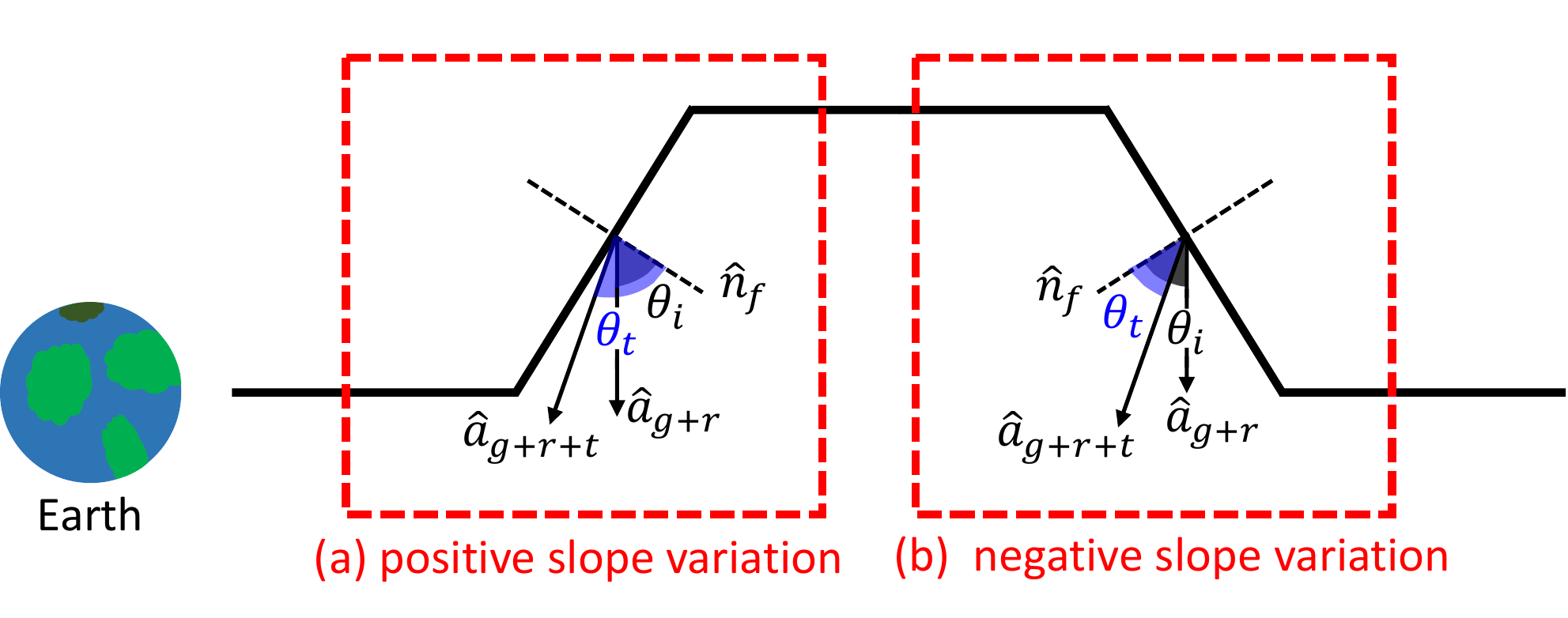}
    \caption{The same initial surface slope has either positive or negative slope variation depending on where the Earth is located in the patch frame. Here, the vectors $\hat{n}_f$,  $\hat{a}_{g+r}$, and $\hat{a}_{g+r+t}$ are the surface normal, the combined self-gravity and rotational acceleration, and the combined $\hat{a}_{g+r}$ and tidally induced acceleration, respectively. The angles $\theta_i$ and $\theta_t$ are the surface slope at the initial and at a specific time, respectively. The positive slope variation is when $\theta_t$ is greater than $\theta_i$ (a), while the opposite case, where $\theta_t$ is smaller than $\theta_i$, is the negative slope variation (b).}
	\label{FIG:temp_slopechanges_patch_shape}
\end{figure}

\begin{figure}
	\includegraphics[width=\columnwidth]{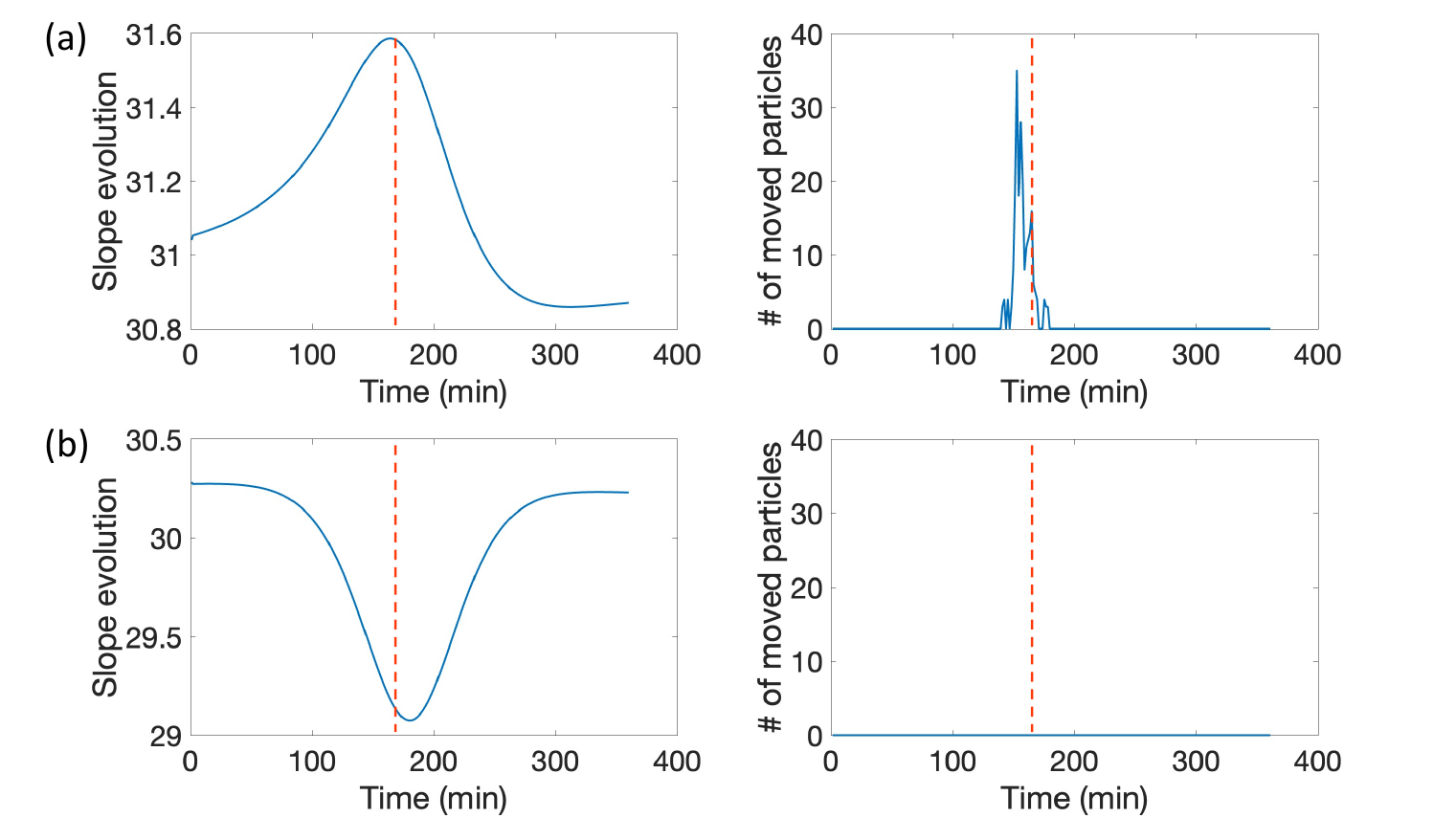}
	\caption{Surface slope evolution and the corresponding particle movements for two patches with similar initial slopes. (a) shows the positive slope variation case when subtracting the peak/trough surface slope from the initial slope for a facet has a positive value}, while (b) shows the negative slope variation case. The left panels show how the surface slope changes over the 6 h Apophis-Earth encounter. The red dotted line marks the time of closest approach. The number of particles exhibiting significant motion is measured at each time step, plotted, and shown in the right-most panels. Note that in the negative slope evolution case (b) no particles move, while the positive slope evolution case (a) does show particle motion.
	\label{FIG:slope_grainmotion}
\end{figure}

As a next step, we convert the generated surface acceleration data into surface slope profile data that includes the initial surface slopes and slope variations for all of the facets on the shape model. We output the total surface acceleration vector for each patch at 60 s intervals to use in the DEM modeling (see Sec.~\ref{Sec:DEMSetup}) and later combine the surface slope profile data with the measured grain motions in an equivalent surface patch in DEM simulations to gain insight into motions across the entire surface (see Sec.~\ref{Sec:results}). The surface slope is defined as the angle between the normal direction of the surface facet and the corresponding surface acceleration vector accounting for self-gravity plus any other accelerations under consideration, such as those due to rotation and tides. The slope variation is computed by subtracting the initial surface slope from the surface slope at a given time during the simulation, and thus can take on positive or negative values. Figure\ \ref{FIG:temp_slopechanges_patch_shape} defines 2 cases with positive and negative slope variations from facets with the same initial surface slope. Depending on where a nearby planetary body (here, Earth) is located, the additional force from the tidal effect can increase the slope (Fig.\ \ref{FIG:temp_slopechanges_patch_shape} (a)) or decrease it (Fig.\ \ref{FIG:temp_slopechanges_patch_shape} (b)). The left panels of Fig.\ \ref{FIG:slope_grainmotion} show examples of the surface slope evolution over time in our models, corresponding to positive (upper left) or negative (lower left) slope variations.


\subsection{DEM model} \label{sec:DEM}
\subsubsection{DEM Code Description}
For the DEM portion of the modeling, we use the parallel \textit{N}-body gravity tree code \textit{pkdgrav} \citep{Richardson2000pkd, Stadel2001pkd}. With \textit{pkdgrav}, we model individual grains in a single surface patch on Apophis as discrete spherical particles that feel interparticle gravitational and contact forces, as well as forces from a uniform gravity field. Particle contacts, including interparticle friction, are modeled with a soft-sphere discrete element method \citep[SSDEM;][]{Schwartz2012ssdem, Sanchez2011ssdem}. SSDEM allows neighboring particles to slightly interpenetrate at the point of contact as a proxy for surface deformation, with the degree of interpenetration mediated by a Hooke's law restoring spring force in the \textit{pkdgrav} implementation, with a linear spring constant representing a material stiffness akin to a Young's modulus \citep{demartini2019using} in the normal direction, plus an equivalent tangential spring component as part of the full spring-dashpot model \citep{Zhang2017demfric}. The SSDEM approach takes user-provided coefficients to account for normal ($\varepsilon_n$) and tangential ($\varepsilon_t$) damping, plus rolling ($\mu_r$), twisting ($\mu_t$), sliding and static ($\mu_s$) friction, and includes a `shape parameter' ($\beta$) to represent grain angularity in rolling interactions \citep{Schwartz2012ssdem, Zhang2017demfric}. This approach has been validated through comparisons with laboratory experiments \citep{Schwartz2013labcomp} and has been used previously by \cite{yu2014numerical} and \cite{demartini2019using} to study potential surface avalanching and bulk reshaping during the Apophis tidal encounter with the Earth.

\subsubsection{DEM Simulation Parameters} \label{Sec:DEMSetup}
The typical facet from the shape model used in the dynamics simulations has mean surface area of 48$\pm$35 m$^2$ (1-$\sigma$). For the SSDEM modeling, we create a single patch with dimensions $(8 \times 8 \times 3)$ m$^3$ in volume, which has a surface area of 64 m$^2$, slightly larger than the mean facet surface area from the shape model but still representative. To create the desired patch of particles for the SSDEM modeling, we settle just over $11,000$ spherical particles in free space under the influence of only interparticle self gravity, with particle radii ($R_p$) ranging from 5.96 to 17.86 cm and with a size-frequency distribution following a power law with slope $\alpha = -3$, roughly matching the size-frequency distributions of decimeter-scale regolith on Bennu's Nightingale Crater \citep{walsh2022packing} and boulders on Itokawa \citep{michikami2008itokawaSFD,mazrouei2014itokawaSFD}. Once the initial cloud of particles has settled into a roughly spherical rubble pile, we carve out a region with periodic lateral boundaries $(8 \times 8 \times 3)$ m$^3$ in volume. We use the same rectangular patch as the initial condition for all of our SSDEM models, as the subsequent tilting stage (described below) provides sufficient randomness in initial particle positions, especially in concert with varying the initial slope and orientation of the patch at the time of encounter from the dynamics models. Throughout the SSDEM modeling, we use friction parameters such that our particle assembly has a friction angle of $\phi = 35.1^\circ$ (see Table~\ref{tab:dem_params}) \citep{lambe2008soil,Zhang2017demfric}, which results in a typical initial patch packing to a bulk density of 2.2 g cm$^{-3}$. In the SSDEM models presented in this study, we do not include the effects of interparticle cohesion, although we aim to investigate this in future work (see Section \ref{Sec:futurework} for details).
    
We continue to prepare the SSDEM patch for each individual encounter by placing an infinite plane below the particles for them to rest on and applying a uniform acceleration normal to the plane surface with magnitude equivalent to the initial pre-encounter acceleration on the given patch from the dynamics model. We choose our patch depth of 3 m such that particles in the upper layers, where resurfacing may occur, will be physically independent of the underlying plane. Over the course of 4 h of simulated time, we rotate the uniform acceleration vector from the normal direction to the actual initial orientation of the acceleration on the patch from the dynamics model. Slowly rotating the uniform acceleration vector is equivalent to quasi-statically tilting the patch to the same initial slope and orientation as used in the dynamics model; we call this the `tilting stage' and we rotate the acceleration vector rather than the particles so that we can remain in the frame of the patch for ease of modeling and visualization. We include an additional 2 h of simulated time after the initial acceleration vector has rotated to its final orientation so that the particles in the patch that have shifted slightly during the tilting stage can reach their equilibrium resting positions.

\begin{table}
	\centering
	\caption{\textit{pkdgrav} DEM Simulation Parameters}
	\label{tab:dem_params}
    \begin{tabular}{p{0.5\columnwidth}p{0.2\columnwidth}p{0.2\columnwidth}}
            \hline
            Quantity & Symbol & Value \\
            \hline
            Particle Number & $N$ & 7141\\
            Particle Radius & $R_p$ & 5.96--17.86 cm\\
            Size-Frequency Distribution Slope & $\alpha$ & --$3.0$\\
            Coefficients of Restitution (*) & $\varepsilon_n, \varepsilon_t$ & 0.55\\
            Coefficient of Static Friction & $\mu_s$ & 1.0\\
            Coefficient of Rolling Friction & $\mu_r$ & 1.05\\
            Coefficient of Twisting Friction & $\mu_t$ & 1.3\\
            Shape Parameter & $\beta$ & 0.7\\
            Angle of Friction & $\phi_f$ & $35.1^\circ$\\
            Initial Patch Bulk Density & $\rho_b$ & 2.2 g cm$^{-3}$\\
            Patch Dimensions (full side lengths) & $l \times w \times h$ & $(8 \times 8 \times 3)$ m$^3$ \\
            \hline
          \multicolumn{3}{p{\columnwidth}}{\textit{Note.} (*) $\varepsilon_n$ and $\varepsilon_t$ define a normal and tangential coefficient, respectively. Spring-dashpot normal and tangential damping coefficients $C_n, C_t$ are dependent on the masses of interacting particles. For any 2-particle interaction: $C_n = C_t \in [9.78, 263]$ kg s$^{-1}$.}
        \end{tabular}
\end{table}

After we have settled a surface patch at the orientation and initial surface acceleration that one of the patches in the dynamics models would feel, we can simulate the full encounter for that patch. We use the tilted and equilibrated patch discussed above and smoothly rotate and change the magnitude of the ambient, uniform acceleration vector in the SSDEM simulations to match the accelerations ($\Vec{a}_{net}$ in Appendix \ref{Appendix_dynamics}) felt by the analogous patch from the dynamics model, interpolated between intervals of 60 s. These simulations last for the same 6 simulated hours as the dynamics models; the particles in the patch uniformly experience the dynamics model accelerations in addition to non-uniform interparticle gravitational and contact accelerations from the other particles. We track particle positions and velocities over time to determine particle motion in the patch and consider each particle that moves by more than half of its radius to exhibit `significant' motion. 

\subsubsection{Resurfacing Estimation} \label{Sec:resurf_calc} 
\begin{figure*}
	\includegraphics[width=2.0\columnwidth]{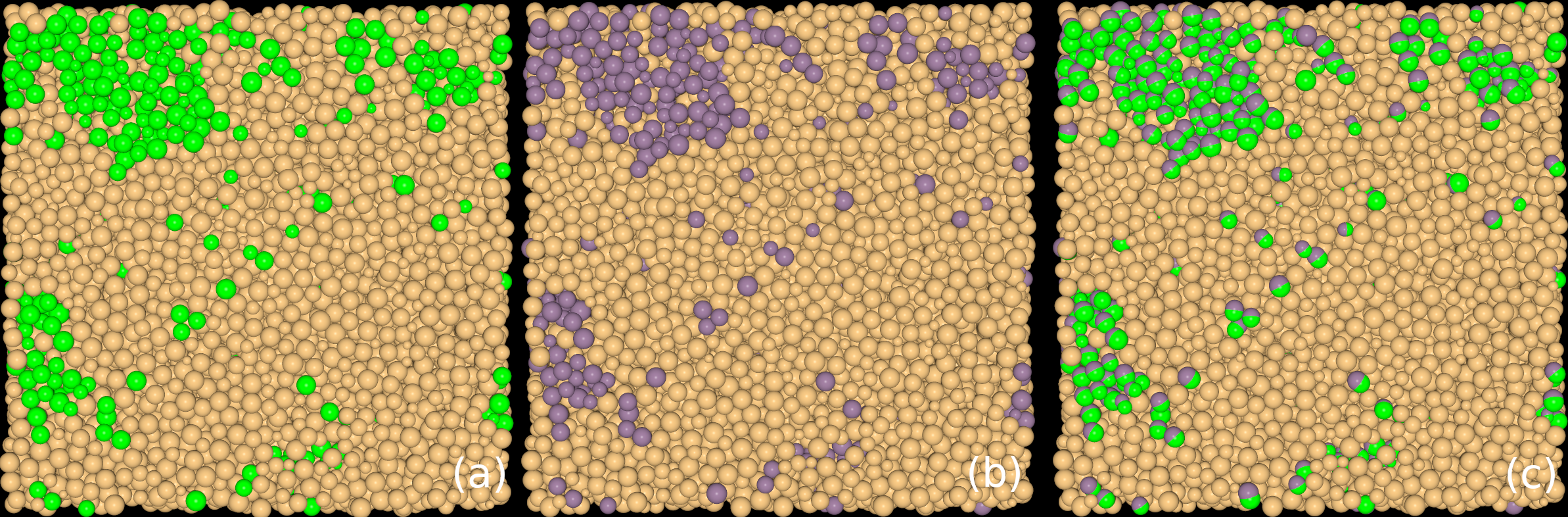}
 	\caption{A depiction of grain motion in a surface patch in the DEM simulations. (a) shows the initial positions of the grains in this patch, with particles that will exhibit `significant motion' (a displacement of more than 1 particle radius) colored green. (b) shows the final positions of particles in the patch, with particles exhibiting significant motion in violet. (c) shows the initial state of the patch, just as in (a), overlaid with the final positions of significantly moved particles, to show downslope motion (from green to violet) toward the upper-left side of the frame. }
 	\label{FIG:DEM_grainmotion}
\end{figure*}

For each particle that has moved significantly in our DEM models, we estimate the amount of revealed unweathered surface area with the following assumptions: 1) each particle exhibits perfect rolling motion directly from its initial to its final position along a straight-line path with no sliding; 2) the half of the particle's surface uncovered from above in the patch frame at the beginning of the encounter is `weathered,' while the remaining half is `unweathered;' 3) the full area initially underneath a moved particle is `unweathered;' and 4) only particle motion in the upper 56 cm (3 times the largest $R_p$) of the particle bed contributes to the patch's resurfacing. We show a sample result from one of our models in Fig.~\ref{FIG:DEM_grainmotion}, where grains exhibiting significant motion are colored green and violet in their initial (left panel) and final (center panel) positions, respectively, with Fig.~\ref{FIG:DEM_grainmotion} (c) showing the final positions (purple) overlaid on the initial patch (green and gold) to help visualize the downslope motion.

\begin{figure}
 		\includegraphics[width=\columnwidth]{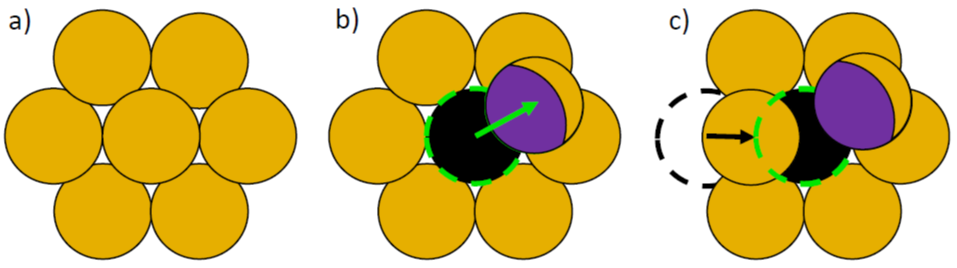}
 	\caption{A schematic diagram indicating how we account for the amount of revealed unweathered area when a particle in the system moves. (a) shows the initial configuration of some particles in the patch. (b) indicates motion of the central particle along the green arrow. The black area with the green dashed outline indicates the revealed unweathered area initially below the moved sphere ($A_u$), while the purple fraction of the moved sphere indicates the revealed area of its surface, which is initially unweathered ($A_{sph}$). (c) indicates another particle moving at a later time, along the path of the black arrow, and covering some of the area that initially counted toward $A_u$ from the central particle's motion. Since some of the black circle has been covered, we no longer count the covered portion toward $A_u$.}
 	\label{FIG:DEM_motionschematic}
 \end{figure}

For a single sphere under the above assumptions, we calculate the total revealed unweathered surface area as the sum of the area of the lens of the sphere's revealed unweathered surface (Equation (\ref{eq:Asph})) plus the fraction of the revealed cross section initially underneath the sphere (Equation~(\ref{eq:Au})):

\begin{equation}\label{eq:Asph}
    A_{sph} = \frac{\pi}{2} R_p^2 \big(1 - \cos{\varphi}\big),
\end{equation}

\begin{equation}\label{eq:Au}
    A_u = \pi R_p^2 \times
    \begin{cases}
        1, & \frac{d}{2R_p} \geq 1 \\
        \frac{d}{2R_p}, & \frac{d}{2R_p} < 1
    \end{cases},
\end{equation}

\noindent where $\varphi$ is the angular displacement of the sphere (assuming perfect rolling motion) and $d$ is the linear displacement of the particle during the simulation. A schematic of this motion is shown in Figs.~\ref{FIG:DEM_motionschematic} (a) and (b), where the central sphere in (a) moves along the green arrow in (b): the revealed unweathered surface area of the sphere ($A_{sph}$) is colored purple, and the cross section of revealed area initially underneath the sphere ($A_u$) is colored black.
    
Under these conditions, a spherical particle with final position exactly one diameter away from its initial position would contribute twice its cross section to the total `resurfaced' area: the full unweathered cross section of the sphere itself plus the full circular cross-sectional area that was initially below the sphere. The total revealed unweathered area in the patch is then calculated as the sum of the area revealed by each sphere, still applying the above assumptions: 
    
\begin{equation}\label{Atot}
    A_{tot} = \sum_{i=0}^N A_{sph,i} + A_{u,i}.
\end{equation}
    
The possibility of one sphere covering an area of surface revealed by another sphere is accounted for by subtracting the area of a lens from the `revealed' area underneath of a sphere based on the initial and final positions of the particles and their relative radii. This scenario is illustrated in Fig.~\ref{FIG:DEM_motionschematic} (c), where another moving sphere covers the black `revealed' area in the center, indicating that we no longer include the newly covered fractional area in $A_u$.

\section{Results}
\label{Sec:results}
Using the tidal resurfacing model, we select 655 surface patches---enough to densely sample the full range of slope variations and initial slopes below the 35$^\circ$ friction angle---and measure the number of grains that exhibit significant motion, as defined above. We sort the simulated patches given their surface slope profiles (initial surface slope and slope variation) and compile the grain motion predictions estimated in the resurfacing estimation phase of the DEM models (Section.~\ref{Sec:resurf_calc}). We then find the correlation between the grain motions and surface slope profiles to constrain the tidal resurfacing across the entire surface (Section \ref{Sub:result1}). In Section \ref{Sub:result2}, we create global surface maps to show the expected resurfaced areas after 3 representative encounter orientations and then demonstrate how the expected resurfaced locations differ depending on the orientation of Apophis at the time of encounter. 

\subsection{Correlation between surface slope profiles and constrained grain motions}
\label{Sub:result1}
\begin{figure}
	\includegraphics[width=\columnwidth]{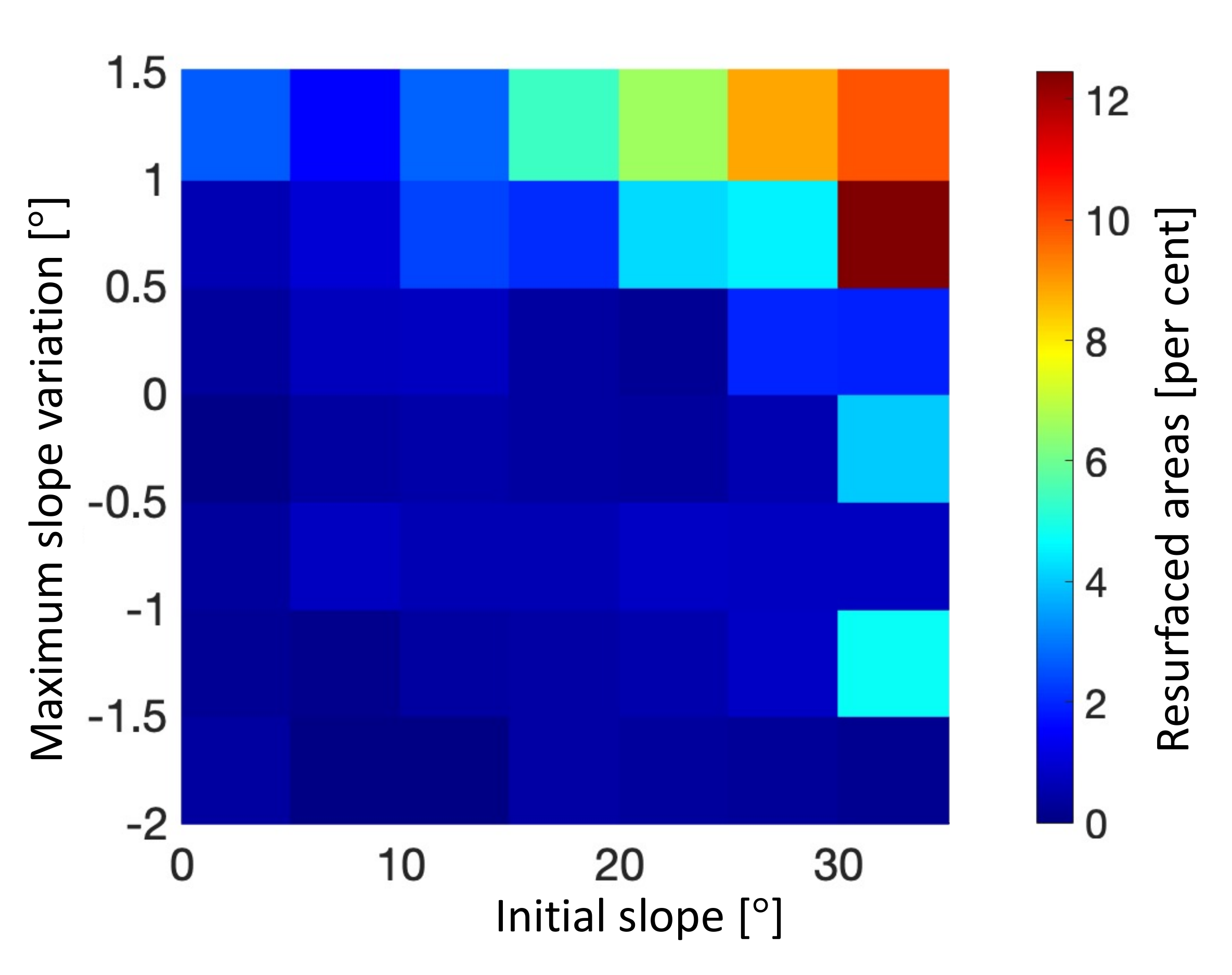}
	\caption{A statistical result showing the correlation between surface slope profiles, namely the initial slope and maximum slope variation, and resurfaced area for each surface patch. Resurfaced area defines the total revealed unweathered surface area in a patch ($A_{tot}$, equation (\ref{Atot})) as a percentage of the entire surface area of the patch (64 m$^2$). Note that the highest initial slope and positive variations give rise to the greatest resurfacing shown in the upper right corner. Areas expected to experience the greatest resurfacing will be the initially high-sloped regions having a positive slope variation, which is affected by the encounter orientation.}
	\label{FIG:StatisticalPlot}
\end{figure}

To discover how the grain motion in a patch correlates with its surface slope profile, we conduct a simple statistical analysis. We first bin the selected 655 surface patches in 2 dimensions: initial slope and slope variation. Here, the bin sizes of initial slope and slope variation are set in increments of 5$^{\circ}$ with a range of $[0^{\circ},\ 35^{\circ}]$ and 0.5$^{\circ}$ with a range of $[-2^{\circ},\ 1.5^{\circ}]$, respectively. On average, each bin includes 13 surface patches. After binning the data, and computing the total resurfaced area for each patch (Section \ref{Sec:resurf_calc}) as a percentage of the total patch area, we compute the average percentage of resurfaced area for the surface patches in each bin. Figure \ref{FIG:StatisticalPlot} shows the average resurfacing in each bin as an `expected' percentage of resurfaced area for a patch given an initial slope and slope variation. 

The first trend we see in Fig.~\ref{FIG:StatisticalPlot} is that the initially high-sloped regions (see the right-hand side of Fig.~\ref{FIG:StatisticalPlot} where the initial slope $> 30^{\circ}$) typically show locally resurfaced areas regardless of the slope variation, which means that the surface grains at high-sloped regions are more susceptible to downslope movement. This trend is supported by the fact that erosion rates by downslope regolith flow become high as surface slopes are close to the angle of repose \citep{culling1960analytical,richardson2014investigating}. The second feature we observe in Fig.~\ref{FIG:StatisticalPlot} is that large slope variation (see the top of Fig.~\ref{FIG:StatisticalPlot}) is a dominating factor in determining regions of resurfacing. We find that most surface patches with slope variations exceeding +$0.5^{\circ}$ show local resurfacing despite some such patches having initial slopes far less than the angle of repose. Granular flow in a region of sub-critical slope (below the angle of repose) has been previously suggested by \cite{ballouz2019surface}, who numerically showed that the tidal forcing from Mars could cause surface mobility on Phobos in areas with significant slope variations, even in regions with slopes less than the angle of repose. Although the tidal forcing on Phobos is different from what Apophis will experience, in that Phobos is continuously under the tidal effects of Mars while Apophis experiences a one-time event from Earth, we still find that the mid-sloped regions (initial slope between $15^{\circ}$ and $30^{\circ}$) on Apophis are likely to experience resurfacing when there is a significant slope variation ($> 0.5^{\circ}$). When grain motions occur in mid-sloped regions, we note that the slope variation increases the patch slope prior to the closest approach distance (positive $y$-axis values in Fig.~\ref{FIG:StatisticalPlot}), while the grains in surface patches with negative slope variations are relatively stable and motionless, despite similar initial slopes and slope variation magnitudes. Figures \ref{FIG:slope_grainmotion} (a) and (b) show how the surface slope evolves (left-most panels) at sampled patches that have a slope increase and decrease, respectively. The corresponding right-side panels of Fig.~\ref{FIG:slope_grainmotion} show the number of particles moving at each timestep measured in the DEM models during the 6-h encounter. The initial slopes for both patches are similarly set to $\sim$30$^{\circ}$. The magnitude of the slope variation is slightly higher in the decreasing case, Fig.~\ref{FIG:slope_grainmotion} (b), but both exceed $0.5^{\circ}$ in absolute magnitude. We observe the surface grains actively moving when the slope variation is positive, however there is no significant grain motion in the patch with negative slope variation. In addition, the most significant grain motion occurs just before perigee when the slope rate of change is highest (the closest encounter happens at 180 min and is marked as a red dotted line in each panel of Fig.~\ref{FIG:slope_grainmotion}). This feature is commonly observed in other patches exhibiting significant grain motions.

\subsection{Influence of encounter orientation on expected resurfaced area}
\label{Sub:result2}
\begin{figure}
	\includegraphics[width=\columnwidth]{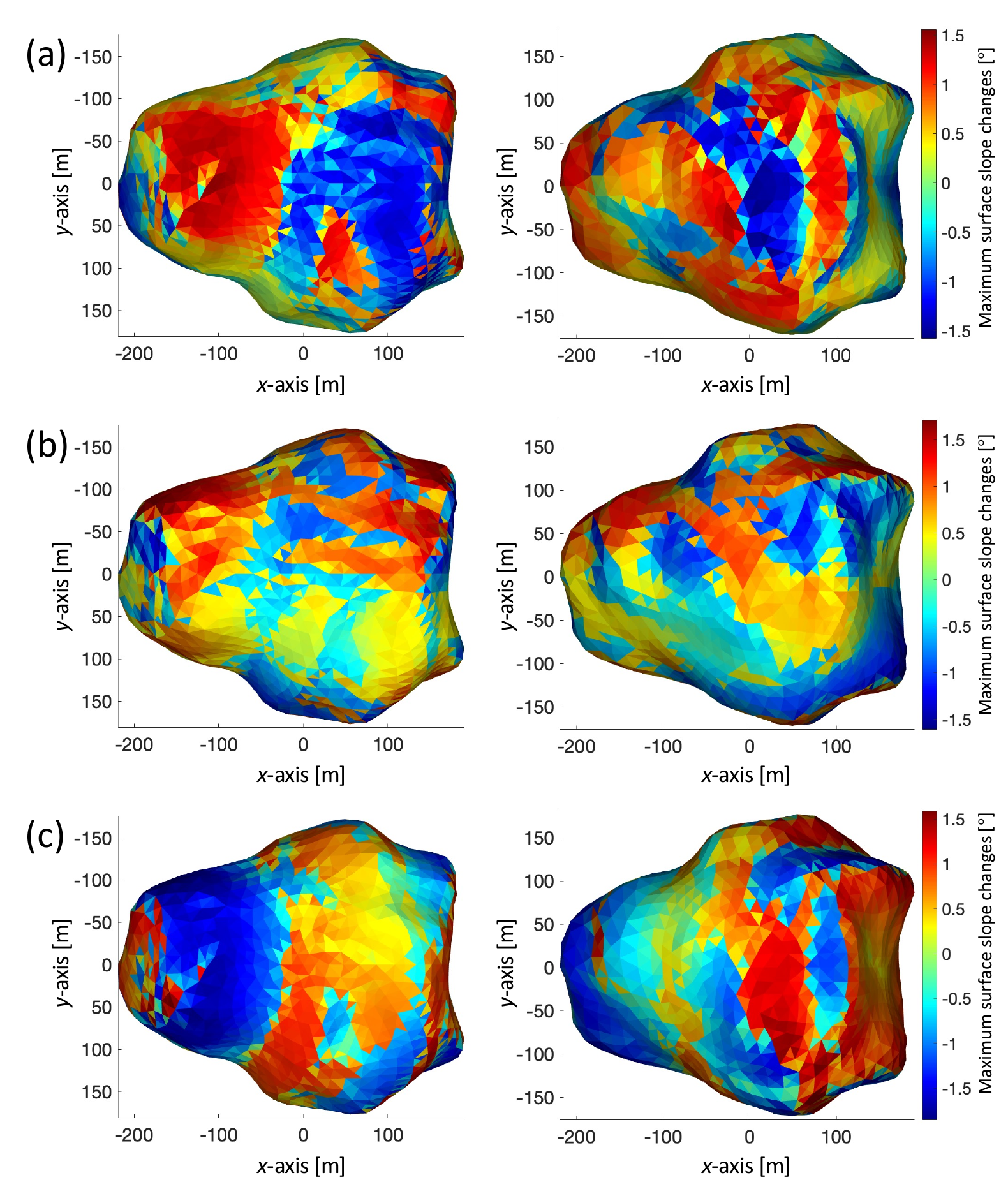}
	\caption{Maximum slope changes with different encounter orientations: the Earth is located above the $x$-$y$ plane (a), the $x$-$z$ plane (b), and the $y$-$z$ plane (c). The left-most maps show the facets above the $x$-$y$ plane, while those on the right-hand side are below the $x$-$y$ plane. All coordinate planes refer to the body-fixed frame of Apophis with the origin at the center of body and $x$- and $z$-axes aligned with the longest and shortest primary body axes, respectively.}
	\label{FIG:MaxSlopeChanges}
\end{figure}

Based on the results from Sec.~\ref{Sub:result1}, we find that the encounter orientation may be a dominant factor in predicting locations and total areas of resurfacing during the 2029 Apophis-Earth close encounter because different encounter orientations cause different surface slope profiles. 
To explore the influence of encounter orientation on regions of tidal resurfacing, we randomly select the encounter orientations rather than propagating from the current spin state because predicting the spin orientation of Apophis at the time of its closest approach to the Earth using the currently existing data still has a large uncertainty \citep{pravec2014tumbling, benson2022spin}. We originally conduct 30 simulations with different encounter orientations but first introduce 3 representative cases chosen to maximize observable differences in resurfacing as a result of encounter orientation. In the dynamics model, we set 3 different initial spin orientations (at a time of 3 h before the closest encounter), which each place Earth above different coordinate planes in the body frame of Apophis at the time of perigee. Figure \ref{FIG:MaxSlopeChanges} shows a surface color map representing the largest magnitude of slope variation (aka.\ Maximum Slope Changes in the colorbar label of Fig.~\ref{FIG:MaxSlopeChanges}) across Apophis's entire surface during the encounter for each orientation. For the encounter orientations, the Earth is located above the $x$-$y$  plane for Fig.~\ref{FIG:MaxSlopeChanges} (a), the $x$-$z$  plane for Fig.~\ref{FIG:MaxSlopeChanges} (b), and the $y$-$z$  plane for Fig.~\ref{FIG:MaxSlopeChanges} (c). 

For all 3 orientation cases, we confirm that the patches showing the largest maximum slope variations are clustered on the side of the object experiencing the strongest tidal forces, where the patches face the Earth for most of the duration of the encounter. As an example, Fig.~\ref{FIG:MaxSlopeChanges} (a), when the Earth is located above the $x$-$y$ plane, shows that the most significant slope variations, both positive and negative, occur for the patches nearest the Earth. Figure~\ref{FIG:MaxSlopeChanges} indicates that the largest slope variations occur in the patches closest to the Earth, but not all patches have positive slope variations because the slope change is affected by the orientation of the patch normal vector compared to the direction of the Earth. Figure \ref{FIG:temp_slopechanges_patch_shape} depicts two cases where the same tidal force induces a positive or negative slope change depending on the different initial orientation of the patches relative to the vector of the Earth's tides. When the tidal force acts along the direction normal to the patch, it induces positive slope variation. The tidal force vector acting opposite the asteroid's gravity prevents grains from resting on the surface and is more likely to induce significant grain motions. In contrast, the force vector causing the negative slope variation plays a role in strengthening the local gravity vector and thus lets grains pack tighter to the surface. This interpretation supports the trend seen in Fig.~\ref{FIG:StatisticalPlot} that more grain motion is observed in patches with positive slope variations rather than negative ones. 

\begin{figure*}
	\includegraphics[width=2.\columnwidth]{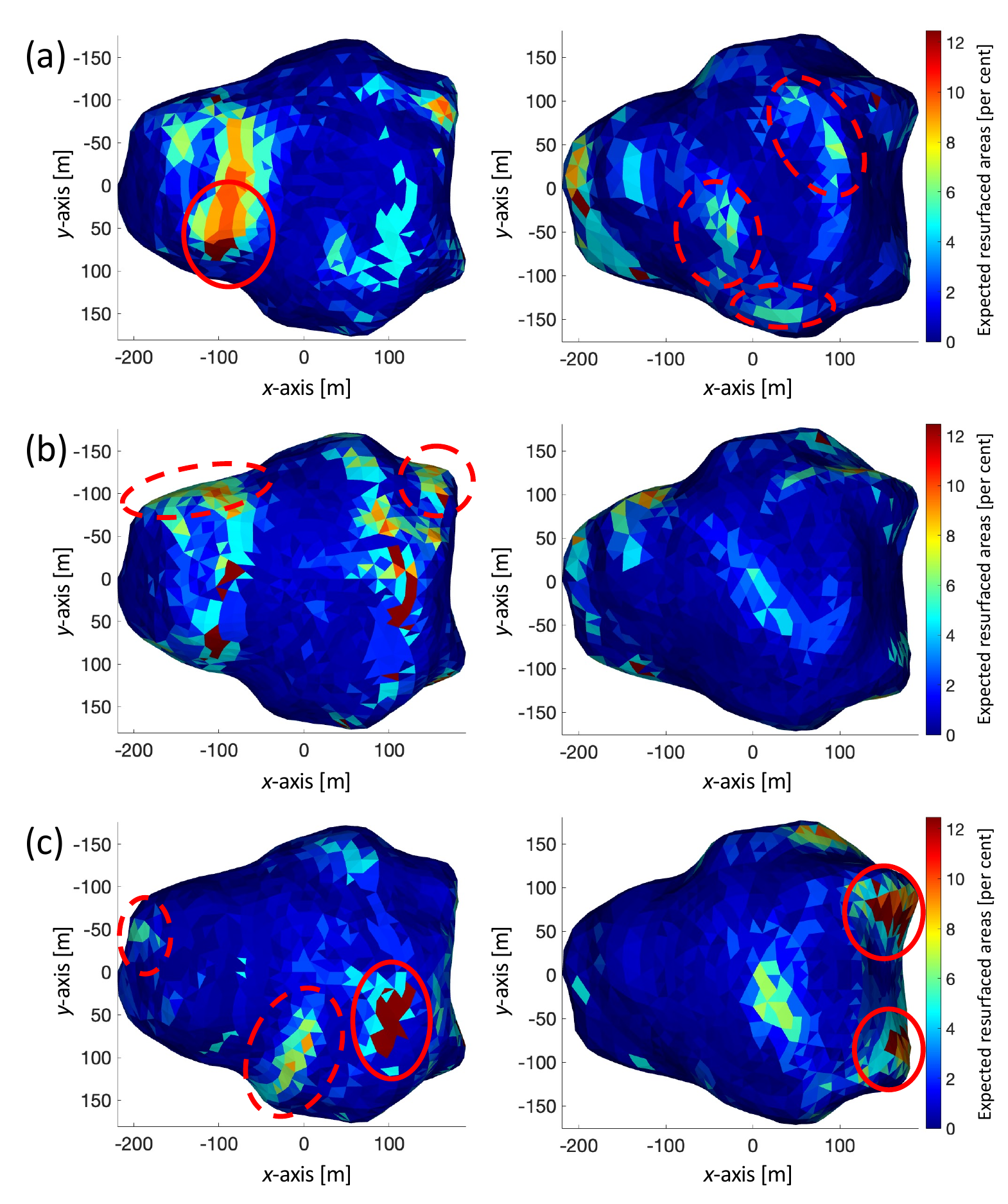}
	\caption{Global surface map showing the expected tidal resurfacing level, which is estimated by the statistical resurfacing result in Fig.~\ref{FIG:StatisticalPlot}, with different encounter orientations: the Earth is located above the $x$-$y$ plane (a), the $x$-$z$ plane (b), and the $y$-$z$ plane (c). The left-most maps show the facets above the $x$-$y$ plane, while those on the right-hand side are below the $x$-$y$ plane. The red solid circles denote locations where the most active grain motions occur at the initially high-sloped regions among 3 orientation cases. The red dotted circles define some locations where our models indicate significant resurfacing only in specific encounter orientations.}
	\label{FIG:Globalresurfacingmap}
\end{figure*}

Using the slope profiles for each encounter orientation driven by the dynamics model, we extrapolate expected areas of resurfacing across the entire surface of Apophis for 3 different encounter orientation cases (see Fig.~\ref{FIG:Globalresurfacingmap}). The amount of resurfacing on each patch is defined by mapping its initial slope and slope variation onto the statistically averaged plot from Fig.~\ref{FIG:StatisticalPlot}. We note that there are common areas that have the resurfaced patches seen in all cases, such as the `neck' region of the contact binary shape. These areas are initially high-sloped regions, with slopes exceeding $30^{\circ}$ or with supercritical initial slopes (higher than the expected friction angle), and are subject to tidal resurfacing regardless of encounter orientation. Despite the expected tendency for initially high-sloped regions to experience resurfacing, we find the encounter orientation still significantly influences how much resurfacing we see at those regions and how much we expect adjacent patches to also experience resurfacing (defining the `width' of the resurfacing region). We mark the common areas that show significant resurfacing and have the widest neighboring resurfaced regions as red solid circles in Fig.~\ref{FIG:Globalresurfacingmap} and find that these regions typically match the initially high-sloped regions with the largest positive slope changes. As an example, the location marked as the red solid circle in Fig.~\ref{FIG:Globalresurfacingmap} (a) shows that significant resurfacing is expected at a high-sloped region on the neck when the Earth is located above the $x$-$y$ plane. However, the same location when the Earth is located above the $y$-$z$ plane (Fig.~\ref{FIG:Globalresurfacingmap} (c)) shows very limited expected tidal refreshing. When we look at the slope changes when Earth is above the $x$-$y$ plane (Fig.~\ref{FIG:MaxSlopeChanges} (a)), we see positive slope variations in the aforementioned area around the neck, while this same region shows decreasing slopes when the Earth is above the $y$-$z$ plane. We reaffirm this trend in another initially high sloped region represented in the red solid circle in Fig.~\ref{FIG:Globalresurfacingmap} (c). At this area, larger positive slope variations occur in the case when the Earth is on the $y$-$z$ plane (Fig.~\ref{FIG:MaxSlopeChanges} (c)) than the other 2 orientation cases (Figs. \ref{FIG:MaxSlopeChanges} (a) and (b)). Besides the commonly resurfaced areas, we notice that there are certain regions that can show strong signals for resurfacing depending on the exact encounter orientation. We mark those regions in dashed red circles in Fig.~\ref{FIG:Globalresurfacingmap}. These regions mostly match with the mid-sloped patches (15 -- 30 deg) nearest the Earth and thus have large slope variations that induce regolith motion. This result again indicates that the regions of expected resurfacing are strongly related to the encounter orientation.

\begin{figure}
	\includegraphics[width=\columnwidth]{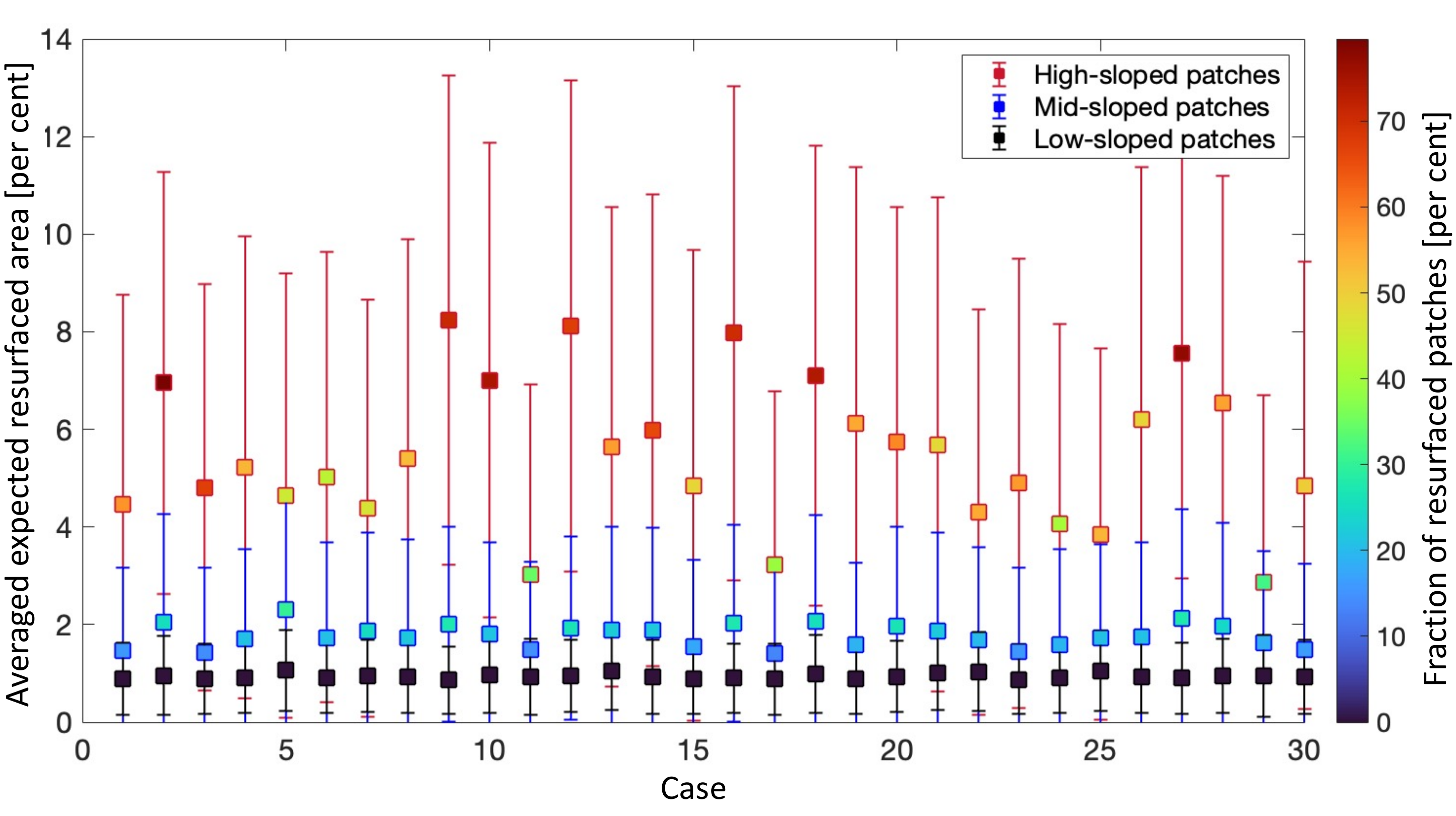}
	\caption{Simulation results for 30 different encounter orientation cases. The red, blue, and black errorbar colors represent the group of patches sorted into the initially high-sloped ($> 30^{\circ}$), mid-sloped ($15^{\circ}$--$30^{\circ}$), and low-sloped ($<  15^{\circ}$) subsets, respectively. The square shows the mean expected resurfaced area for each group, with error bars representing one standard deviation in the positive or negative directions. The color of the square represents the fraction of patches in that subset that have a total expected resurfaced area of at least 3 per cent of the total patch area.}
	\label{FIG:30_EO_cases}
\end{figure}

To support the features we found by comparing 3 representative encounter orientation cases, we statistically investigate the estimated resurfacing for high-, mid-, and low-sloped patch subsets in 30 encounter orientation cases (see Fig.~\ref{FIG:30_EO_cases}). Our metric for quantifying how many patches in each subset are being significantly resurfaced is the number of patches for which $A_{tot} \geq 0.03A_{patch}$. The mean expected resurfaced area in mid- and low-sloped patches is less than 3\%, thus the value (0.03) is chosen in the above inequality to most clearly delineate the color difference in the fraction of resurfaced patches among low-, mid-, and high-sloped patches. As expected, the high-sloped patches have the largest fraction experiencing significant resurfacing for all encounter orientations---represented by the colorbar in Fig.~\ref{FIG:30_EO_cases}. The results indicate a maximum of 80 per cent and a minimum of 32 per cent of all initially high-sloped patches have more than 3 per cent of their total area resurfaced across our 30 orientation simulations. In all cases, this group has a higher level of average expected resurfaced area and larger standard deviations than the initially mid- and low-sloped patch subsets, despite variations dependent on the encounter orientation. The trend here supports the feature indicated in our 3 representative cases: that the degree of resurfacing and the width of the resurfacing regions at the initially high-sloped locations depend strongly on the encounter orientation. For the mid-sloped patch subsets, our results show that a small fraction ($\sim$10 per cent) of the patches experience significant resurfacing, and have a lower expected resurfaced area than the higher-sloped patches. As we confirmed in the comparison of our 3 representative encounter orientations, the resurfaced regions in the mid-sloped subset of patches match areas with significant positive slope variations. Unlike the high-sloped and mid-sloped patches, we confirm that the low-sloped patch subset is stable against tidal resurfacing regardless of the encounter orientation. As a final note, we address that all of our simulated cases indicate very local resurfacing, with a total expected resurfaced area of only 1 per cent of the entire Apophis surface area.

\section{Discussion}
\label{Sec:Discussions}
Our simulation results indicate that the initially high-sloped regions show more grain motion than the low-sloped regions with a similar slope variation. This finding indicates that initially high-sloped regions are more sensitive to tidal refreshing because even a relatively small slope variation can make the surface slope exceed the friction angle. This result may indicate potential common areas that will experience tidal resurfacing during the Earth flyby regardless of the encounter orientation. If the initially high-sloped regions from the current shape model truly exist on Apophis itself, however, these areas might already exhibit fresher and brighter regolith than other areas before the encounter. The high-resolution images from in-situ missions to S-types Itokawa and Eros show evidence of surface color variations on both bodies. Earlier studies (e.g., \cite{miyamoto2007regolith, gaskell2008characterizing}) found a correlation between these surface color variations and the surface slope distribution: the higher-sloped areas tend to show brighter surfaces because of down-slope grain motion that exposed fresher material underneath the weathered top layer, while the lower-sloped areas were covered in loose, weathered regolith believed to have migrated there. In consideration of the features seen on the surfaces of Itokawa and Eros, the initially high-sloped regions on the Sq-type Apophis's surface might already show fresher and brighter regolith unrelated to the upcoming tidal interaction, or as a result of previous tidal resurfacing. If we could obtain a surface map of the pre-encounter state of Apophis from a time when the tidal effect of the 2029 Earth encounter is negligible, comparing it against a similar post-encounter surface map would be a way to accurately confirm whether the upcoming planetary encounter will cause any brighter surface colors at the initially high-sloped regions. Besides that, we also note that some initially high-sloped regions from the shape model may be not be realistic, given that the current Apophis shape model still has significant uncertainties, as addressed by \cite{brozovic2018goldstone}.

We also find that the encounter orientation is the dominating factor in predicting more targeted areas where we could detect tidal refreshing, given that a positive slope variation is more likely to induce surface grain motion. Since the same location in the body frame can have positive or negative slope variation depending on where the Earth is located in the patch frame, any area can be subject to resurfacing. When the Earth's tides are close to being aligned with the patch normal or downslope direction (Fig.~\ref{FIG:temp_slopechanges_patch_shape} (a)), the tidal force competes with local gravity, preventing the grains from resting on the surface, thereby inducing motion. This means that we can predict areas of expected tidal refreshing on the surface of Apophis given more accurate knowledge of the encounter orientation at perigee. Currently, the most feasible way to predict the encounter orientation of Apophis during the 2029 flyby is by propagating the well-constrained spin state from the 2012--2013 apparition to the 2029 apparition while considering the potential tidal effects from the Earth, non-principal-axis rotation, and minor effects of Yarkovsky and YORP that could dynamically alter the object's rotational state. Unfortunately, there are still significant uncertainties in the spin-state data from past apparitions \citep{pravec2014tumbling, brozovic2018goldstone}, and the computational errors associated with propagating the spin state over $\sim$16 yr are fairly significant. Given these uncertainties, the best way to improve our knowledge of the 2029 encounter orientation must come from ground observations in the time just before the 2029 encounter: the DSS-13 and DSS-14 antennas at Goldstone will start in mid-March and the DSS-43 antenna at the Canberra Deep Space Communication Complex in Australia will cover the time around the closest approach \citep{brozovic2022radar}. We anticipate that these pre-encounter radar observations will provide accurate detail about the rotational state of Apophis, which can then be used to constrain the likely areas of resurfacing, so that those areas can be targeted for confirmation by potential spacecraft and ground-based observations. 

Lastly, we confirm that active grain motion most commonly occurs in the half hour before perigee (as shown in the right-side frame of Fig. \ref{FIG:slope_grainmotion} (a)) and with total resurfacing on the scale of $\sim$1 per cent of Apophis's entire surface area. These findings provide essential information about the timing and scale of potential tidal resurfacing, which could support the mission planning of observation campaigns and OSIRIS-APEX. In most patches, surface grains move actively when the patch slopes change most rapidly; the changing surface accelerations can make the grains unstable and move from their equilibrium positions. This means that the best time to detect active tidal resurfacing is during the last hour before Apophis reaches perigee, after which the surface grains reach new equilibrium positions because the rates of change in the patch slopes become smaller. We also note that the global tidal resurfacing that we predict is not extensive, which is consistent with conclusions from previous studies \citep{yu2014numerical, benson2022spin}. However, our results still indicate that tidal resurfacing may be seen in certain localized regions---initially high-sloped areas and mid-sloped regions with significant positive slope variation. Given these indications, ground-based observations could detect the level of tidal resurfacing that our models predict if there are precise surface images or albedo maps both before and after the closest encounter. \cite{brozovic2022radar} address that Goldstone DSS-13 and DSS-14 antennas can obtain high-resolution radar images that would place tens of thousands of pixels on Apophis during the time 10 days before to 10 days after the closest encounter. That kind of radar observation covers the time of the 6-h encounter considered in our simulations and can help provide a new, very detailed shape model, possibly capturing surface features as small as a few meters in size. Not only the radar observations, but optical observations can also support the shape refinement and obtain a database of Apophis's surface albedos (the optical telescopes that could observe Apophis are listed in Table 2 in \cite{vallejo2022conditions}). The new model prior to the encounter can be used for more refined dynamics modeling to better constrain resurfaced areas. If tidal resurfacing indeed occurs as we predict, we expect evidence of tidal resurfacing to be detectable via analysis of surface images or albedo changes at the resurfaced areas predicted by the refined dynamics model. Considering the small scale of tidal resurfacing we predict, a change in the moments of inertia due to tidal resurfacing is likely to be minimal. Surface properties such as roughness and grain size distributions, important factors in our DEM modeling, could also be better characterized by dual-polarization imaging. Furthermore, other radar facilities such as the 10 GHz HUSIR (Haystack Ultrawideband Satellite Imaging Radar in Westford, Massachusetts) can resolve finer surface features with image resolutions down to a few centimeters near the time of perigee, which could be enough to detect active tidal refreshing. Lastly, the expected surface images from the OSIRIS-APEX mission can be combined with ground-based observations to better understand the influence of the tidal encounter on the surface, despite the spacecraft arriving at Apophis 4 months after the close encounter \citep{dellagiustina2022osiris, nolan2022osiris}.

\section{Future work}
\label{Sec:futurework}
We address two main limitations for the current tidal resurfacing model in the DEM stage: 1) the relatively low porosity of the patch; and 2) the neglecting of cohesive forces in our models. The initial patch used for the DEM modeling stages has a $\sim$45 per cent porosity as a result of allowing the particles to coalesce under self gravity before carving out the shape of our patch (here we are referring to macroporosity between grains, i.e., 45 per cent porosity implies 55 per cent of the volume is occupied by solid grains and the remaining 45 per cent is void space between the grains). For comparison, the estimated bulk porosity for both of the rubble piles Bennu and Ryugu, recently visited by the OSIRIS-REx and Hayabusa2 spacecrafts, respectively, is around 50 per cent \citep{lauretta2019bennusurf, watanabe2019ryugumain}, while the surface regolith layer is estimated to have a significantly higher porosity \citep{walsh2022packing}. This indicates that we are underestimating the porosity of the patches in our DEM models and likely underestimating the degree of resurfacing. Creating systems with significantly different porosities (by more than a few percent) would require preparing the patch in a way that is ad hoc or possibly unphysical, using unrealistic material parameters, adding cohesion during the packing stage, or modeling with irregular particle shapes \citep[e.g.,][]{demartini2022porosity, marohnic2022aggs}. The more void space there is in the patch, the more easily the patch can shear, as the structures maintaining the particle configuration will be less stable. The sample acquisition from the OSIRIS-REx mission to Bennu gave results that indicate potentially 70 per cent or higher porosity in the upper regolith layer on Bennu \citep{walsh2022packing}, implying a more `fluffy' surface regolith structure than we are modeling here. Apophis is an Sq-type asteroid, which, based on comparisons with Sq-type Itokawa \citep{barnouin2008itokawafines} and S-type Eros \citep{cheng2002erosfines}, may imply the presence of more fine-grained surface material compared to the boulder-heavy surface of the B-type Bennu \citep{lauretta2019bennusurf}. Since the porosity approximation comes from Bennu's Nightingale Crater, where there are more fines than other regions of the surface \citep{barnouin2022nightingale}, and without significant additional data about Sq-type surfaces, we still believe Bennu to be a satisfactory point of comparison. Regardless, we intend to investigate the effects of modeling regolith systems with higher porosities in a future study.

In contention with the low porosity of our patch, which may be reducing the amount of resurfacing we see, is the absence of cohesion in our models. Interparticle cohesive forces like Van der Waals forces can increase the shear strength of a granular assembly, keeping the system stable against the relatively weak tidal forces felt by any given surface patch. At the scale we model, where particles are tens of centimeters in diameter, we expect very low interparticle cohesion, if any \citep{sanchez2014cohesion}. Still, even a small amount of cohesion could be enough to restrict the resurfacing that we see in our models. This will be investigated in our future work.

\section{Conclusions}
This study visits the topic of tidal refreshing on the surface of Apophis, a phenomenon that may be observable during the 2029 Apophis-Earth close encounter, using our tidal resurfacing model. The main finding in this work is that the tidal resurfacing likely occurs at small scales in very localized regions, mostly 30 min before the closest encounter. In particular, the orientation of Apophis at the time of closest approach will control which areas on the surface may experience tidally induced resurfacing. If detailed surface topographic maps or albedo data during the encounter event can be obtained through the collaboration of ground-based observation campaigns and in-situ missions, we may detect evidence supporting local tidal resurfacing as a result of the close Earth encounter.

\section{Data availability}
The data underlying this article will be shared on reasonable request to the corresponding author.

\section*{Acknowledgements}
This work is partly supported by Auburn University’s Intramural Grant Program,  by Zonta Amelia Earhart Fellowship, by FINESST Award 80NSSC21K1531, and by NASA SSERVI Grant 80NSSC19M0216. Y. Kim led the dynamics modeling part, and J. DeMartini was tasked with DEM simulations carried out at the University of Maryland on the Deepthought2 supercomputing cluster administered by the Division of Informational Technology. The dynamics modeling makes use of the SPICE Toolkits for MATLAB. The analysis uses MATLAB and Python’s NumPy and SciPy modules.



\bibliographystyle{mnras}
\bibliography{example} 




\appendix

\section{Dynamics model equations}\label{Appendix_dynamics}
In our dynamics model, the net acceleration vector acting on an Earth-crossing asteroid, $\Vec{a}_{net}$, is defined as:
\begin{equation}\label{Eq_acc}
        \Vec{a}_{net} = \Vec{a}_{g} + \Vec{a}_{t} + \Vec{a}_{r},
    \end{equation}
where $\Vec{a}_{g}$ is a self-gravity term, $\Vec{a}_{t}$ is a tidal force term, and $\Vec{a}_{r}$ is a rotational force term. Each term can be written as:
\begin{equation}\label{Eq_acc_g}
        \Vec{a}_{g} = G\rho \int_{V_A}\frac{\Vec{r}}{r^3} dV,
    \end{equation}
    
\begin{equation}\label{Eq_acc_t}
        \Vec{a}_{t} = \frac{GM_E}{{R_c}^3}(\Vec{r}-\frac{3({\Vec{R}_c}+\Vec{r})\cdot\Vec{r}}{R_c^2}{\Vec{R}}_c),
    \end{equation}
    
\begin{equation}\label{Eq_acc_r}
        \Vec{a}_{r} = \dot{\vec{{\omega}}}\times\vec{r} + \Vec{\omega}\times\dot{\Vec{r}} + \Vec{\omega}\times(\Vec{\omega}\times\Vec{r}),
    \end{equation}
where $G$ is the gravitational constant, $\rho$ is the object's bulk density, $V$ is the object's volume, $\Vec{r}$ is the position of a surface facet relative to the center of mass of the object, $M_E$ is the mass of the Earth, $\vec{R_c}$ is the position of the Earth relative to the center of mass of the object, $\Vec{\omega}$ is the object's angular velocity vector, $\dot{\Vec{\omega}}$ is the object's angular acceleration vector, and $\dot{\Vec{r}}$ is the time-rate of change in $\Vec{r}$. For $\Vec{a}_{g}$, we follow the approach by \cite{werner1994gravitational} to compute the gravitational potential of the object, represented as a polyhedral mesh. For $\Vec{a}_{t}$, the trajectory of Apophis is retrieved from SPICE kernels \citep{acton2018look} and then converted into a vector in the body-fixed frame of the object to determine $\vec{R_c}$. In the conversion, the transformation matrix ($[A]$) representing Apophis' orientation toward the Earth is updated in the spin state propagation (described in equation (\ref{Eq:rot2})). All constant parameters are defined as the values in the JPL Solar System Dynamics database\footnote{JPL Solar System Dynamics Website (\url{https://ssd.jpl.nasa.gov/planets/phys_par.html})} in addition to $\rho$ = 2 g cm$^{-3}$, which is consistent with Sq-type Itokawa \citep{abe2006mass}. For $\Vec{a}_{r}$, we note that the second term representing the Coriolis effect is ignored, given that the tidally induced shape deformation will be minimal within the current perigee \citep{demartini2019using}. The time-varying $\Vec{\omega}$, $\dot{\Vec{\omega}}$, and $[A]$ are propagated according to the following equations (Euler's rotation equations) with a fourth-order Runge-Kutta integrator:
\begin{equation}
[I]\Dot{\Vec{\omega}} + \Vec{\omega}\times[I]\Vec{\omega} = \frac{3GM_E}{R^5_c}\Vec{R_c}\times[I]\Vec{R_c}, \label{Eq:rot1}
\end{equation}
\begin{equation}
[\dot{A}] = -[\Tilde{{\omega}}][A], \label{Eq:rot2}
\end{equation}
where $[I]$ is the object's moment of inertia and $[\Tilde{{\omega}}]$ is the skew matrix of $\vec{{\omega}}$ defined as below:
\begin{equation}
\Tilde{\omega} = \begin{bmatrix}
0 & -\omega_z & \omega_y \\
\omega_z & 0 & -\omega_x \\
-\omega_y & \omega_x & 0
\end{bmatrix}.
\label{Eq:rot3}
\end{equation}
In the spin state propagation, the initial condition of $\omega_z$ is set as 30.6 h, while other components ($\omega_x$ and $\omega_y$) are zero. For the 3 representative orientation simulations, [A] is initialized to set a given coordinate axis in the body-fixed frame of Apophis to point toward the Earth.


\bsp	
\label{lastpage}
\end{document}